\documentclass[sigconf]{acmart}
\usepackage{amsmath}
\AtBeginDocument{%
  \providecommand\BibTeX{{%
    \normalfont B\kern-0.5em{\scshape i\kern-0.25em b}\kern-0.8em\TeX}}}

\setcopyright{acmcopyright}
\copyrightyear{2022}
\acmYear{2022}
\acmDOI{XXXXXXX.XXXXXXX}

\acmConference[WWW'23]{The Web Conference 2023}{April 01-- May 04, 2022}{Austin, USA}

\begin{document}

\title{Unified Vision-Language Representation Modeling for E-Commerce Same-style Products Retrieval}

\author{Ben Chen, Linbo Jin, Xinxin Wang, Dehong Gao*, Wen Jiang, Wei Ning}

\affiliation{%
  \institution{Alibaba Group}
  \streetaddress{699 Wang Shang Road}
  \city{Hangzhou}
  \state{Zhejiang}
  \country{China}
  \postcode{43017-6221}
}

\email{{chenben.cb, yuyi.jlb, rooney.wxx, dehong.gdh, wen.jiangw, wei.ningw}@alibaba-inc.com}

\renewcommand{\shortauthors}{Ben et al.}

\begin{abstract}
 Same-style products retrieval plays an important role in e-commerce platforms, aiming to identify 
 the same products which may have different text descriptions or images. It can be used for similar products retrieval from different suppliers or duplicate products detection of one supplier. 
 Common methods use the image as the detected object, but they only consider the visual features and overlook the attribute information contained in the textual descriptions, and perform weakly for products in image less important industries like machinery, hardware tools and electronic component, even if an additional text matching module is added.
 In this paper, we propose a unified vision-language modeling method for e-commerce same-style products 
 retrieval, which is designed to represent one product with its textual descriptions and visual contents. 
 It contains one sampling skill to collect positive pairs from user
 click logs with category and relevance constrained, and a novel contrastive loss unit to model the image, text, 
 and image+text representations into one joint embedding space. 
 It is capable of cross-modal product-to-product retrieval, as well as style transfer and user-interactive search. 
 Offline evaluations on annotated data demonstrate its superior retrieval performance, and online testings
 show it can attract more clicks and conversions. Moreover, this model has already been deployed online for 
 similar products retrieval in alibaba.com, the largest B2B e-commerce platform in the world.
\end{abstract}



\maketitle

\section{Introduction}
\textbf{S}ame-\textbf{S}tyle \textbf{P}roducts \textbf{R}etrieval (SSPR) is an important task in e-commerce search system, with the aim to identify products which are the same but released by suppliers with different images and titles. 
It is an essential requirement for the similar products retrieval and duplicate products detection.
Specifically, buyers want to view more of the same products from different suppliers, 
so that they can find the favorite with lower price and better quality, and we should provide this access 
when they are searching. While for suppliers, they tend to deliver same products for hacking more traffic,
so an efficient duplicate products detection skill is in need to suppress this behavior for better users' 
experience and platform health.

\begin{figure}[tb]
  \centering
  \includegraphics[width=\linewidth]{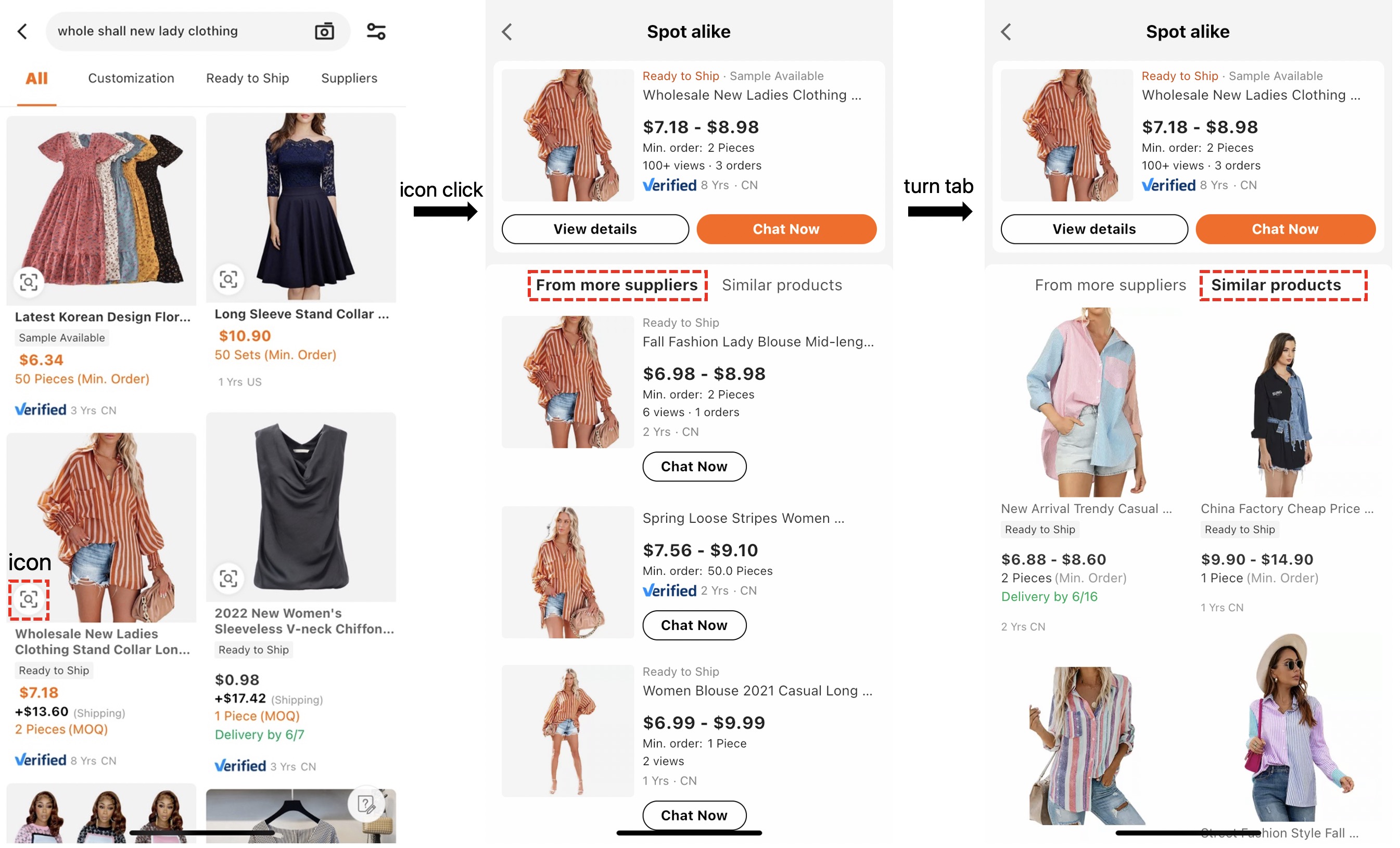}
  \caption{The overview of the similar products retrieval platform in alibaba.com, where "From more suppliers" means the same-style products from different suppliers.}
  \Description{The overview of similar products retrieval platform, where products are recalled based on the learned embedding provided by proposed method.}
  \label{fig:figure1}
\end{figure}

Existing methods for SSPR mainly consider the visual attributes such as product shape, style and color.
They usually adopt an image-to-image matching model to score the images similarity of all candidate 
products, and then select the ones whose score is greater than a certain threshold as the same products. 
But these image-based methods overlook the textual attributes (specific mode, material, brand e.t.c.) 
contained in the textual descriptions, and perform weakly for products in image less important industries, 
e.g. machinery, hardware tools and electronic component. Some supplementary measures will adopt an additional text-to-text matching model to further filter products with irrelevant attributes. However, suppliers may add 
some unrelated keywords into the title for more exposures, and these keywords will mislead the model to make 
wrong judgments. Therefore, to accurately distinguish the same products, we should 
equally consider the image and textual descriptions, and utilize both visual and textual features to model
the product.

In recent years, more and more vision-language pre-training (VLP) models \cite{su2019vl, NEURIPS2019_c74d97b0, 10.1007/978-3-030-58577-8_8, li2020unicoder, gao2020fashionbert,li2022blip}
have emerged to explore how to align the representations of different modals. They try to map the textual and 
visual features into one joint embedding space based on the self-attention mechanism \cite{NIPS2017_3f5ee243, devlin-etal-2019-bert}, and achieve impressive performance in many cross-modal tasks like text-image retrieval, 
image captioning, visual question answering. However, These VLP models mainly focus on text and image matching 
but pay less attention to the image+text to image+text matching for SSPR, of which the key challenge 
is the alignment of image, text and image+text presentations to reduces interference of irrelevant attributes.


To accurately represent one product with its textual description and visual content, this paper propose an unified vision-language representation method. It consists of three types of contrastive loss, with the aim to reduce the representation discrepancy of the product image, text, and image+text from different perspectives. The integrated objective function unit combining these losses can effectively model these three representations into a joint embedding space, and make the key features involved in image
and text dominate the generation of embeddings. 
This method is applied in the fine-tune stage, and can easily adapt to various VLP models with less burden. In addition, In order to collect enough data to train an effective model, we also propose an intuitive training data sampling technique, which can gather high reliable positive pairs from user click logs by constraining categories and relevances.


\begin{figure}[tb]
  \centering
  \includegraphics[width=\linewidth]{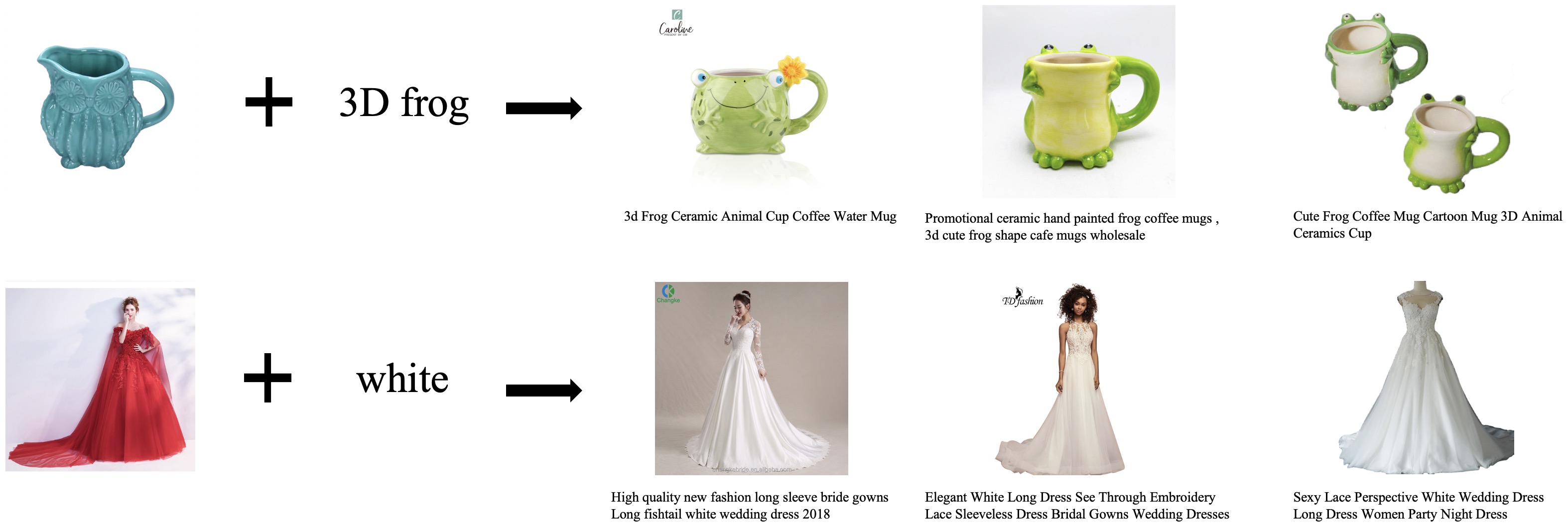}
  \caption{User-interactive search cases with proposed model. A mug with owl design + "3D frog" 
  words can search the frog shape mugs, and a red dress + "white" get the white dresses.}
  \Description{A red dress + "white" word can get corresponding white dresses, shows the potential for use-interactive search.}
  \label{fig:figure2}
\end{figure}

We execute extensive offline evaluations on annotated same-style data for product (image, text, image+text) to product (image, text, image + text) retrieval, and the significant performance boosts demonstrate the proposed method's effectiveness for cross-modal representation. Online tests also prove that it can improve the same-style products coverage rate, and attract more clicks and conversions. This method has already been deployed online for similar products retrieval in alibaba.com, the largest B2B e-commerce platform in the world, as seen in Figure \ref{fig:figure1}.
Moreover, it also shows the potential for applications like style transfer and user-interactive search (shown in Figure \ref{fig:figure2}). 



\section{Method}

\subsection{Query-Item Click Graph}
For training an effective model to retrieval the same-style products, one primary shortcoming is the 
lack of high-reliable training data. Because the manual annotation samples are accurate but 
expensive, while online available data is noisy and can not cover all categories. Here we collect 
reliable samples from user query-item click log, and impose category and relevance restrictions to 
ensure the accuracy of data. This sampling skill is based on a basic cognition that the query reflect 
user's intents and their click behaviors indicates their attention to different products. Two samples 
with deeper click level in the same query should have higher similarity. 

In alibaba.com, the user's click behavior can be divided into 5 levels: page-click ($c$) $\leq$ add-to-cart ($a$) 
$\leq$ contact-supplier ($s$) $\leq$ order ($o$) $\leq$ pay ($p$). We construct the query-item click graph with query nodes 
containing the search texts and the item nodes composed of <image, title, keywords, brand>. Then we use 
the sum of different click level scores as weight to build a co-clicking edge between query $q_i$ and 
item node $c_i$ pairs. Click level score is represented as a coefficient $\boldsymbol{\lambda}$
multiply the click number and final weight is: 
\begin{equation}
  Weight = \lambda_1 cnt_c+\lambda_2 cnt_a+\lambda_3 cnt_s+\lambda_4 cnt_o+\lambda_5 cnt_p.
\end{equation}
For deeper clicks, this coefficient should be set larger to alleviate the effect of <$q_i$, $c_i$> pairs which users click by mistake. 

\subsection{Constrained Graph Sampling}
After constructing the click graph, we use the weighted sampling strategy to generate training samples. 
Specifically, we collect no more than 4 items of one query according to the weight from large to small, 
and build positive sample in pairs. In addition, we set three constraints to ensure the sampling accuracy:

\paragraph{\textbf{Query contains at least two words and one core keyword}} 
Some users have no clear search 
intents and will enter query with ambiguous semantics, e.g. dress, sport shoes. These queries cover a 
relatively broad category and the items clicked by users are also scattered. So we select the query with at
least two words, and one core keyword that clearly represents the user's needs.

\paragraph{\textbf{Two items must be of same sub-category}}
E-commerce platform has a hierarchical and fine-grained category structure, and one category contains dozens of sub-categories. For example, "woman dress" contains 20 sub-categories like "evening dress", "career dress", "casual dress", and "casual dress". A query "red long dress with v-neck" will retrieval items cross these categories. To collect same-style ones, we should keep two items of same sub-category.

\paragraph{\textbf{Image-text pairs should have high similarity}}
Deep click levels limitation can not entirely guarantee that two items are of same-style. So we compute image similarity with embedding outputted from Resnet50 \cite{he2016deep} and text similarity with embedding outputted from Bert
\cite{devlin-etal-2019-bert} for each pairs, and eliminate those pairs with image similarity or text similarity values lower than 0.7.


After the constrained sampling, we select those pairs that contain at least one core keyword in common as the final training samples. This operation is also designed to reduce the sampling error.

\subsection{Contrastive Loss Unit}
The basic architecture of vision-language pre-training (VLP) models is usually composed of a visual embedding module, a textual embedding module and a fusion encoder \cite{yu2021mixbert, yu2022commercemm}. Of these architectures the image is usually encoded with an off-the-shelf ResNet \cite{he2016deep}, or Faster-RCNN \cite{ren2015faster}, or Visual Transformer \cite{dosovitskiy2020image} model. While textual embeddings are generated by dividing the text description into a sequence of word tokens and then fed into one transformer model \cite{devlin-etal-2019-bert, conneau2019unsupervised, liu2019roberta}. Then an encoder will fuse both embeddings to create a contextual interacted multimodal representation.

As for the training objectives, common VLP models use image+text contrastive loss (ITC) to align the embedding 
space of text and image, while adopt binary classification image+text matching loss (ITM) for learning a multimodal representation to predict whether image and text is in pair, as well as some additional objectives like language modeling loss to enhance the embedding learning \cite{gao2020fashionbert, radford2021learning, li2022blip, li2021align}. However, they pay less attention to the feature alignment at the item (image+text) level.

Here we further consider how to generate one multimodal representation used for image+text to 
image+text matching in same-style products retrieval, and proposed a novel contrastive loss unit acting in the fine-tune stage. 
It contains three types of objectives: product to product matching losses (\textbf{PPM}), product self-distinctiveness losses (\textbf{PDC}) and product locality consistency losses (\textbf{PLC}). 
They are designed to learn a cross-modal aligned representation from different perspectives, and all contribute to model image, text, and image+text into one joint embedding space.


\begin{figure}[tb]
  \centering
  \includegraphics[width=\linewidth]{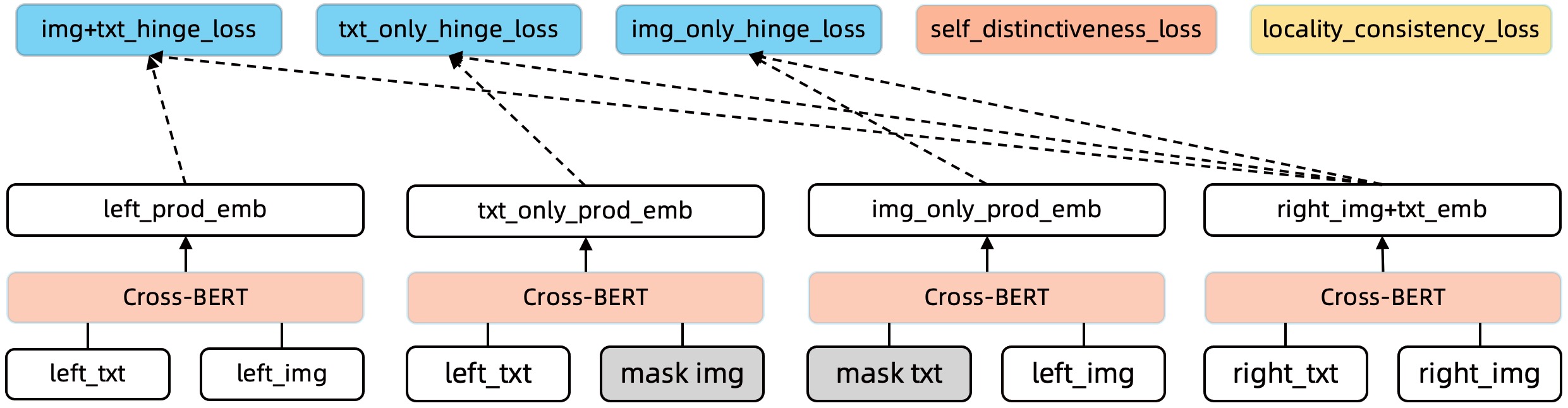}
  \caption{The overview of proposed contrastive loss unit. We mainly show the PPM construction method for brevity.}
  \Description{The proposed contrastive loss unit acts in the fine-tune stage, which models the product text, image, and image+text representation in one joint embedding space.}
\end{figure}

Given an item-item pair of same-style, we take the left item as trigger item $t_i$, and the right item as recall item $r_i$, where the subscript $i$ indicates the $i$-th pair in dataset. 
The multimodal image+text 
embeddings \textbf{$t^m_i$}, \textbf{$r^m_i$} are the average of all embeddings of the last layer.
The single image visual embeddings \textbf{$t^v_i$}, \textbf{$r^v_i$} and textual embeddings \textbf{$t^t_i$}, \textbf{$r^t_i$} are the average of corresponding embeddings of the last layer, with the masked text description and image are replaced with the all-zero sequences to simulate single modal inputs.
Finally, all these embeddings are normalized. 

\paragraph{\textbf{Product to product matching losses (PPM)}} This objective aims to align the representations
of trigger item and recall item into one joint embedding space. We assume that if two items are of same-style, the multimodal image+text embeddings of one should have the high similarity to the other. 
Furthermore, the image, text description (text, keywords, brand, e.t.c) of the trigger item should also be aligned to
the multimodal embeddings of the recall item, making the key features involved in image and text
dominate the generation of embeddings.

Specifically, during fine-tuning stage, each items <$t_i$, $r_i$> of same-style are taken as positive sample pair, while the trigger item $t_i$ and recall item  $r_j$ from other trigger item in the same batch combine the negative sample pair <$t_i$, $r_j$>. We compute the similarity score between three embeddings of trigger item and image+text embeddings of the recall item as the follows:
\begin{equation}
s^{m \leftrightarrow m}_{ij} = \textbf{$t^m_i$} \cdot \textbf{$r^m_j$},  \quad
s^{v \leftrightarrow m}_{ij} = \textbf{$t^v_i$} \cdot \textbf{$r^m_j$},  \quad
s^{t \leftrightarrow m}_{ij} = \textbf{$t^t_i$} \cdot \textbf{$r^m_j$}.
\end{equation}

Then we use hinge loss to guide item pairs of same-style to have greater similarity values compared with those of different styles:
\begin{equation}
\begin{aligned}
L_{PPM} = \frac{1}{N^2} \underset{i}\sum \underset{j}\sum &\frac{1}{3} [max(0, \alpha_1 (1-y_{ij}) + s^{m \leftrightarrow m}_{ij} -s^{m \leftrightarrow m}_{ii}) \\
&+ max(0, \alpha_1 (1-y_{ij}) +s^{v \leftrightarrow m}_{ij} -s^{v \leftrightarrow m}_{ii}) \\
&+ max(0, \alpha_1 (1-y_{ij}) +s^{t \leftrightarrow m}_{ij}-s^{t \leftrightarrow m}_{ii})
],
\end{aligned}
\end{equation}
where $N$ is the batch size, $\alpha_1$ is the margin value. It lets cross-modal embeddings of positive pairs be closer than those of negative pairs.

\paragraph{\textbf{Product self-distinctiveness losses (PDC)}} PDC is designed to ensure that three modal embeddings of one item should have the smaller distance than the multimodal embeddings between trigger item and recall item. In other words, the similarity values between \textbf{$t^v_i$}, \textbf{$t^t_i$} and \textbf{$t^m_i$} should larger than those between \textbf{$t^m_i$} and \textbf{$r^m_j$}.
\begin{equation}
\begin{aligned}
L_{PDC} = \frac{1}{N^2} \underset{i}\sum \underset{j}\sum &\frac{1}{2} [max(0, \alpha_2+ s^{m \leftrightarrow m}_{ij} -s^{v \leftrightarrow m}_{ii}) \\
&+ max(0, \alpha_2 +s^{m \leftrightarrow m}_{ij}-s^{t \leftrightarrow m}_{ii})
],
\end{aligned}
\end{equation}
where $\alpha_2$ is the margin value. This objective is similar to PPM, but it is one effective way to reduce representation discrepancy of three modals from another perspective.

\paragraph{\textbf{Product locality consistency losses (PLC)}} PLC tries to align the embedding of image, text, and image+text to reduces interference of irrelevant attributes. That is, three normalized embeddings of one trigger item should be as consistent as possible. It can be described as the mean square error (MSE) of the similarity values $s^{m \leftrightarrow m}_{ij}$, $s^{v \leftrightarrow m}_{ij}$, and $s^{t \leftrightarrow m}_{ij}$ should smaller than a certain value:
\begin{equation}
\begin{aligned}
L_{PLC} = \frac{1}{N^2} \underset{i}\sum \underset{j}\sum &\frac{1}{3} [max(0, -\alpha_3 + (s^{v \leftrightarrow m}_{ij} - s^{m \leftrightarrow m}_{ij})^2) \\
&+ max(0, -\alpha_3 + (s^{t \leftrightarrow m}_{ij} - s^{m \leftrightarrow m}_{ij})^2) \\
&+ max(0, -\alpha_3 + (s^{v \leftrightarrow m}_{ij} - s^{t \leftrightarrow m}_{ij})^2) 
],
\end{aligned}
\end{equation}
where $\alpha_3$ is the margin value. This computation is relatively large, and we can compute only top 10 pairs for each trigger item, where the rank list is arranged according to $s^{t \leftrightarrow m}_{ij}$ from large to small.

So the total loss is formulated as follows:
\begin{equation}
\begin{aligned}
L_{total} = \frac{1}{3} (L_{PPM} + L_{PDC} + L_{PLC}),
\end{aligned}
\end{equation}

As a side note, the pairs in the same batch should keep the same sub-category but belong to different same-style. This operation drives the model to focus on hard samples, and learn the fine-grained difference between styles. 

\section {Experiments}
In this section, we conduct comprehensive evaluations on annotated same-style data offline and A/B test online to verify the feasibility of proposed method.

\textbf{Dataset.} We collect a large scale of training samples from user click log by the sampling skills detailed above, with the coefficient $\boldsymbol{\lambda}$ set as (1,2,2,5,5). It contains 4 million trigger-recall item pairs of 20 industries and 4000 sub-categories. Since there is no public evaluation data, here we collect 5196 manually annotated pairs covering 15 industries and 973 sub-categories for the offline testing. 

\textbf{Implementation details.} We use tiny FashionBert (L2-H128-A6) \cite{gao2020fashionbert} as the base pre-trained model. It is an economical and efficient model and have been online applied in the item recall stage of many search platforms. The image is split into $4 \times 4$ patches, and passes through ResNet-50 to generate 16 tokens. The maximum text sequence length is set to 50 with the special [CLS] on the head, so the total token length is 66. Learning rate is set as 3e-4, the batch size is 256 and the margin values ${\alpha_{1,2,3}}$ is set as (0.3, 0.2, $0.05^2$). All the evaluated models are trained within 5 epochs and we adopt the early-stopping strategy.

\subsection{\textbf{Offline evaluation.}} 
For the fair comparison with same model structure and parameters, we fine-tune two baseline models with only image (base v-v) and text (base t-t) as the inputs while the corresponding objectives are set to be image-image and text-text hinge loss, then use the outputs (visual embeddings, textual embeddings) of trigger and recall items to compute the similarities. We abbreviate the proposed model as eSSPR for short. eSSPR v-v / t-t / m-m denote that we fine-tune pre-trained model with image + text as the inputs and $L_{total}$ as the loss function, and we compute the similarities with only image, only text and image+text as the inputs to generate corresponding embeddings.
For all evaluations we adopt Mean Reciprocal Ranking (MRR) and recall@K as the metrics, which are widely used in search and recommendation systems. All data presented are the average values of the metrics for all testing pairs.

\begin{table}
  \caption{Comparison with different methods on manual annotated data. R@n is the abbreviation of Recall@n}.
  \label{tab:offline}
  \begin{tabular}{lccccc}
    \toprule   
    Method & MRR & R@1 & R@5 & R@10 & R@20 \\
    \midrule	  
    base t-t & 0.5454 & 0.4744 & 0.6407 & 0.6938 & 0.7404 \\
    eSSPR t-t & 0.6428 & 0.5597 & 0.7485 & 0.8174 & 0.8761 \\
     \midrule	  
    base v-v & 0.8620 & 0.8389 & 0.8905 & 0.9121 & 0.9284 \\
    eSSPR v-v & 0.9027 & 0.8784 & 0.9355 & 0.9511 & 0.9675 \\
     \midrule	  
    eSSPR m-m & \textbf{0.9197} & \textbf{0.8870} & \textbf{0.9619} & \textbf{0.9761} & \textbf{0.9879} \\
   \bottomrule   
\end{tabular}
\end{table}

\begin{table}
  \caption{Ablation experiment results of different losses with same inputs. R@n is the abbreviation of Recall@n}.
  \label{tab:ablation}
  \begin{tabular}{lccccc}
    \toprule   
    Loss & MRR & R@1 & R@5 & R@10 & R@20 \\
    \midrule	  
    $L_{base}$\cite{gao2020fashionbert} & 0.8915 & 0.8582 & 0.9342 & 0.9542 & 0.9654 \\
    $L_{PPM}$ & 0.8935 & 0.8543 & 0.9463 & 0.9646 & 0.9779 \\
    $L_{PPM+PDC}$ & 0.9065 & 0.8730 & 0.9524 & 0.9695 & 0.9801 \\
    $L_{PPM+PLC}$ & 0.9131 & 0.8784 & 0.9596 & \textbf{0.9761} & 0.9852 \\
     \midrule	  
    $L_{total}$ & \textbf{0.9197} & \textbf{0.8870} & \textbf{0.9619} & \textbf{0.9761} & \textbf{0.9879} \\
   \bottomrule   
\end{tabular}
\end{table}

The results are shown in Table~\ref{tab:offline}, we can see that eSSPR m-m outperforms two baseline method with large improvements, demonstrating the effectiveness of the proposed unified vision-language representation method. Furthermore, even if we only use unimodal features (text or image), the model eSSPR t-t (v-v) achieves a higher performance than base t-t (v-v), which proves that the contrastive loss unit contributes to make the key features involved in image and text dominate the generation of embeddings. 

In order to verify the feasibility of proposed loss unit, we provide additional ablation experiments, which are exhibited in Table~\ref{tab:ablation}. The $L_{base}$ is the hinge loss of image+text embeddings for trigger items (\textbf{$t^m_i$}) and recall items (\textbf{$r^m_i$}). We find that $L_{PPM}$ have the better performance than $L_{base}$. Combined with the results of unimodal models  in Table~\ref{tab:offline}, we believe that $L_{PPM}$ can make the representation of image, text and image+text mutually enhance each other. 
$L_{PPM+PLC}$ and $L_{PPM+PDC}$ improve MRR and recall rate significantly, and  $L_{total}$ achieves the best results among them.
These comparisons verify the contrastive loss unit is capable of boosting the performance substantially.

\subsection{\textbf{Online Experiments.}} 
We lauched the proposed model in our similar products retrieval platform, which is used in the search, homepage recommender and you-shopping-history recommender modules in alibaba.com (shown in Figure \ref{fig:figure1}). We compare the experimental performance of SCR (same-style products coverage rate), CTR (click through rate) and CVR (average conversion rate) before and after the launch. As shown in Table~\ref{tab:onlinex}, the SCR, CTR and CVR metrics increase $+64.9\%$, $+2.51\%$, $+2.31\%$ respectively. The results verified that the proposed unified vision-language representation method can attract more clicks and conversions for increasing the industry revenue.

\begin{table} [htbp]
  \caption{Online results for A/B testing}.
  \label{tab:onlinex}
  \begin{tabular}{ccc}
    \toprule   
    SCR & CTR & CVR \\
    \midrule	  
    +64.9\% & +2.51\% & +2.31\% \\
    \bottomrule   
\end{tabular}
\end{table}

More cross-modal retrieval results, the annotated testing data cases, corresponding codes and the application demo of proposed methods in alibaba.com are attached in the appendix. All of them will be released to public through the official repository, as well as the adaptive testings for modern dual-stream models like CLIP \cite{radford2021learning} and Align \cite{li2021align}.

\section{Conclusion}
Same-style products retrieval (SSPR) provides one channel for users to compare more same products with lower price or higher quality. To identify the same products which may have different text descriptions and images, we proposed a unified vision-language representation method for same-style products retrieval. It contains a reliable training samples construction technique, and a novel contrastive loss unit acts in the fine-tune stage to model the product image, text, and image+text representation into one joint embedding space. Extensive experiments on offline annotated item pairs demonstrates its superior performance for the same products retrieval. Online testings have verified that it can attract more clicks and conversions in the similar product recommendation in alibaba.com. In the further, we will focus on its extended applications likes style transfer and user-interactive search, and also study how to make the more fine-grained alignment of multimodal features for e-commerce platform.

\bibliographystyle{ACM-Reference-Format}
\bibliography{sample-cikm}


\begin{thebibliography}{17}


\ifx \showCODEN    \undefined \def \showCODEN     #1{\unskip}     \fi
\ifx \showDOI      \undefined \def \showDOI       #1{#1}\fi
\ifx \showISBNx    \undefined \def \showISBNx     #1{\unskip}     \fi
\ifx \showISBNxiii \undefined \def \showISBNxiii  #1{\unskip}     \fi
\ifx \showISSN     \undefined \def \showISSN      #1{\unskip}     \fi
\ifx \showLCCN     \undefined \def \showLCCN      #1{\unskip}     \fi
\ifx \shownote     \undefined \def \shownote      #1{#1}          \fi
\ifx \showarticletitle \undefined \def \showarticletitle #1{#1}   \fi
\ifx \showURL      \undefined \def \showURL       {\relax}        \fi
\providecommand\bibfield[2]{#2}
\providecommand\bibinfo[2]{#2}
\providecommand\natexlab[1]{#1}
\providecommand\showeprint[2][]{arXiv:#2}

\bibitem[Conneau et~al\mbox{.}(2019)]%
        {conneau2019unsupervised}
\bibfield{author}{\bibinfo{person}{Alexis Conneau}, \bibinfo{person}{Kartikay
  Khandelwal}, \bibinfo{person}{Naman Goyal}, \bibinfo{person}{Vishrav
  Chaudhary}, \bibinfo{person}{Guillaume Wenzek}, \bibinfo{person}{Francisco
  Guzm{\'a}n}, \bibinfo{person}{Edouard Grave}, \bibinfo{person}{Myle Ott},
  \bibinfo{person}{Luke Zettlemoyer}, {and} \bibinfo{person}{Veselin
  Stoyanov}.} \bibinfo{year}{2019}\natexlab{}.
\newblock \showarticletitle{Unsupervised cross-lingual representation learning
  at scale}.
\newblock \bibinfo{journal}{\emph{arXiv preprint arXiv:1911.02116}}
  (\bibinfo{year}{2019}).
\newblock


\bibitem[Devlin et~al\mbox{.}(2019)]%
        {devlin-etal-2019-bert}
\bibfield{author}{\bibinfo{person}{Jacob Devlin}, \bibinfo{person}{Ming-Wei
  Chang}, \bibinfo{person}{Kenton Lee}, {and} \bibinfo{person}{Kristina
  Toutanova}.} \bibinfo{year}{2019}\natexlab{}.
\newblock \showarticletitle{{BERT}: Pre-training of Deep Bidirectional
  Transformers for Language Understanding}. In
  \bibinfo{booktitle}{\emph{Proceedings of the 2019 Conference of the North
  {A}merican Chapter of the Association for Computational Linguistics: Human
  Language Technologies}}. \bibinfo{publisher}{Association for Computational
  Linguistics}, \bibinfo{address}{Minneapolis, Minnesota},
  \bibinfo{pages}{4171--4186}.
\newblock


\bibitem[Dosovitskiy et~al\mbox{.}(2020)]%
        {dosovitskiy2020image}
\bibfield{author}{\bibinfo{person}{Alexey Dosovitskiy}, \bibinfo{person}{Lucas
  Beyer}, \bibinfo{person}{Alexander Kolesnikov}, \bibinfo{person}{Dirk
  Weissenborn}, \bibinfo{person}{Xiaohua Zhai}, \bibinfo{person}{Thomas
  Unterthiner}, \bibinfo{person}{Mostafa Dehghani}, \bibinfo{person}{Matthias
  Minderer}, \bibinfo{person}{Georg Heigold}, \bibinfo{person}{Sylvain Gelly},
  {et~al\mbox{.}}} \bibinfo{year}{2020}\natexlab{}.
\newblock \showarticletitle{An image is worth 16x16 words: Transformers for
  image recognition at scale}.
\newblock \bibinfo{journal}{\emph{arXiv preprint arXiv:2010.11929}}
  (\bibinfo{year}{2020}).
\newblock


\bibitem[Gao et~al\mbox{.}(2020)]%
        {gao2020fashionbert}
\bibfield{author}{\bibinfo{person}{Dehong Gao}, \bibinfo{person}{Linbo Jin},
  \bibinfo{person}{Ben Chen}, \bibinfo{person}{Minghui Qiu},
  \bibinfo{person}{Peng Li}, \bibinfo{person}{Yi Wei}, \bibinfo{person}{Yi Hu},
  {and} \bibinfo{person}{Hao Wang}.} \bibinfo{year}{2020}\natexlab{}.
\newblock \showarticletitle{Fashionbert: Text and image matching with adaptive
  loss for cross-modal retrieval}. In \bibinfo{booktitle}{\emph{Proceedings of
  the 43rd International ACM SIGIR Conference on Research and Development in
  Information Retrieval}}. \bibinfo{pages}{2251--2260}.
\newblock


\bibitem[He et~al\mbox{.}(2016)]%
        {he2016deep}
\bibfield{author}{\bibinfo{person}{Kaiming He}, \bibinfo{person}{Xiangyu
  Zhang}, \bibinfo{person}{Shaoqing Ren}, {and} \bibinfo{person}{Jian Sun}.}
  \bibinfo{year}{2016}\natexlab{}.
\newblock \showarticletitle{Deep residual learning for image recognition}. In
  \bibinfo{booktitle}{\emph{Proceedings of the IEEE conference on computer
  vision and pattern recognition}}. \bibinfo{pages}{770--778}.
\newblock


\bibitem[Li et~al\mbox{.}(2020a)]%
        {li2020unicoder}
\bibfield{author}{\bibinfo{person}{Gen Li}, \bibinfo{person}{Nan Duan},
  \bibinfo{person}{Yuejian Fang}, \bibinfo{person}{Ming Gong}, {and}
  \bibinfo{person}{Daxin Jiang}.} \bibinfo{year}{2020}\natexlab{a}.
\newblock \showarticletitle{Unicoder-vl: A universal encoder for vision and
  language by cross-modal pre-training}. In
  \bibinfo{booktitle}{\emph{Proceedings of the AAAI Conference on Artificial
  Intelligence}}, Vol.~\bibinfo{volume}{34}. \bibinfo{pages}{11336--11344}.
\newblock


\bibitem[Li et~al\mbox{.}(2022)]%
        {li2022blip}
\bibfield{author}{\bibinfo{person}{Junnan Li}, \bibinfo{person}{Dongxu Li},
  \bibinfo{person}{Caiming Xiong}, {and} \bibinfo{person}{Steven Hoi}.}
  \bibinfo{year}{2022}\natexlab{}.
\newblock \showarticletitle{Blip: Bootstrapping language-image pre-training for
  unified vision-language understanding and generation}.
\newblock \bibinfo{journal}{\emph{arXiv preprint arXiv:2201.12086}}
  (\bibinfo{year}{2022}).
\newblock


\bibitem[Li et~al\mbox{.}(2021)]%
        {li2021align}
\bibfield{author}{\bibinfo{person}{Junnan Li}, \bibinfo{person}{Ramprasaath
  Selvaraju}, \bibinfo{person}{Akhilesh Gotmare}, \bibinfo{person}{Shafiq
  Joty}, \bibinfo{person}{Caiming Xiong}, {and} \bibinfo{person}{Steven
  Chu~Hong Hoi}.} \bibinfo{year}{2021}\natexlab{}.
\newblock \showarticletitle{Align before fuse: Vision and language
  representation learning with momentum distillation}.
\newblock \bibinfo{journal}{\emph{Advances in Neural Information Processing
  Systems}} (\bibinfo{year}{2021}).
\newblock


\bibitem[Li et~al\mbox{.}(2020b)]%
        {10.1007/978-3-030-58577-8_8}
\bibfield{author}{\bibinfo{person}{Xiujun Li}, \bibinfo{person}{Xi Yin},
  \bibinfo{person}{Chunyuan Li}, \bibinfo{person}{Pengchuan Zhang},
  \bibinfo{person}{Xiaowei Hu}, \bibinfo{person}{Lei Zhang},
  \bibinfo{person}{Lijuan Wang}, \bibinfo{person}{Houdong Hu},
  \bibinfo{person}{Li Dong}, \bibinfo{person}{Furu Wei}, \bibinfo{person}{Yejin
  Choi}, {and} \bibinfo{person}{Jianfeng Gao}.}
  \bibinfo{year}{2020}\natexlab{b}.
\newblock \showarticletitle{Oscar: Object-Semantics Aligned Pre-training for
  Vision-Language Tasks}. In \bibinfo{booktitle}{\emph{Computer Vision -- ECCV
  2020}}. \bibinfo{publisher}{Springer International Publishing},
  \bibinfo{address}{Cham}, \bibinfo{pages}{121--137}.
\newblock


\bibitem[Liu et~al\mbox{.}(2019)]%
        {liu2019roberta}
\bibfield{author}{\bibinfo{person}{Yinhan Liu}, \bibinfo{person}{Myle Ott},
  \bibinfo{person}{Naman Goyal}, \bibinfo{person}{Jingfei Du},
  \bibinfo{person}{Mandar Joshi}, \bibinfo{person}{Danqi Chen},
  \bibinfo{person}{Omer Levy}, \bibinfo{person}{Mike Lewis},
  \bibinfo{person}{Luke Zettlemoyer}, {and} \bibinfo{person}{Veselin
  Stoyanov}.} \bibinfo{year}{2019}\natexlab{}.
\newblock \showarticletitle{Roberta: A robustly optimized bert pretraining
  approach}.
\newblock \bibinfo{journal}{\emph{arXiv preprint arXiv:1907.11692}}
  (\bibinfo{year}{2019}).
\newblock


\bibitem[Lu et~al\mbox{.}(2019)]%
        {NEURIPS2019_c74d97b0}
\bibfield{author}{\bibinfo{person}{Jiasen Lu}, \bibinfo{person}{Dhruv Batra},
  \bibinfo{person}{Devi Parikh}, {and} \bibinfo{person}{Stefan Lee}.}
  \bibinfo{year}{2019}\natexlab{}.
\newblock \showarticletitle{ViLBERT: Pretraining Task-Agnostic Visiolinguistic
  Representations for Vision-and-Language Tasks}. In
  \bibinfo{booktitle}{\emph{Advances in Neural Information Processing
  Systems}}, Vol.~\bibinfo{volume}{32}. \bibinfo{publisher}{Curran Associates,
  Inc.}
\newblock


\bibitem[Radford et~al\mbox{.}(2021)]%
        {radford2021learning}
\bibfield{author}{\bibinfo{person}{Alec Radford}, \bibinfo{person}{Jong~Wook
  Kim}, \bibinfo{person}{Chris Hallacy}, \bibinfo{person}{Aditya Ramesh},
  \bibinfo{person}{Gabriel Goh}, \bibinfo{person}{Sandhini Agarwal},
  \bibinfo{person}{Girish Sastry}, \bibinfo{person}{Amanda Askell},
  \bibinfo{person}{Pamela Mishkin}, \bibinfo{person}{Jack Clark},
  {et~al\mbox{.}}} \bibinfo{year}{2021}\natexlab{}.
\newblock \showarticletitle{Learning transferable visual models from natural
  language supervision}. In \bibinfo{booktitle}{\emph{International Conference
  on Machine Learning}}. PMLR, \bibinfo{pages}{8748--8763}.
\newblock


\bibitem[Ren et~al\mbox{.}(2015)]%
        {ren2015faster}
\bibfield{author}{\bibinfo{person}{Shaoqing Ren}, \bibinfo{person}{Kaiming He},
  \bibinfo{person}{Ross Girshick}, {and} \bibinfo{person}{Jian Sun}.}
  \bibinfo{year}{2015}\natexlab{}.
\newblock \showarticletitle{Faster r-cnn: Towards real-time object detection
  with region proposal networks}.
\newblock \bibinfo{journal}{\emph{Advances in neural information processing
  systems}}  \bibinfo{volume}{28} (\bibinfo{year}{2015}).
\newblock


\bibitem[Su et~al\mbox{.}(2019)]%
        {su2019vl}
\bibfield{author}{\bibinfo{person}{Weijie Su}, \bibinfo{person}{Xizhou Zhu},
  \bibinfo{person}{Yue Cao}, \bibinfo{person}{Bin Li}, \bibinfo{person}{Lewei
  Lu}, \bibinfo{person}{Furu Wei}, {and} \bibinfo{person}{Jifeng Dai}.}
  \bibinfo{year}{2019}\natexlab{}.
\newblock \showarticletitle{Vl-bert: Pre-training of generic visual-linguistic
  representations}.
\newblock \bibinfo{journal}{\emph{arXiv preprint arXiv:1908.08530}}
  (\bibinfo{year}{2019}).
\newblock


\bibitem[Vaswani et~al\mbox{.}(2017)]%
        {NIPS2017_3f5ee243}
\bibfield{author}{\bibinfo{person}{Ashish Vaswani}, \bibinfo{person}{Noam
  Shazeer}, \bibinfo{person}{Niki Parmar}, \bibinfo{person}{Jakob Uszkoreit},
  \bibinfo{person}{Llion Jones}, \bibinfo{person}{Aidan~N Gomez},
  \bibinfo{person}{\L~ukasz Kaiser}, {and} \bibinfo{person}{Illia Polosukhin}.}
  \bibinfo{year}{2017}\natexlab{}.
\newblock \showarticletitle{Attention is All you Need}. In
  \bibinfo{booktitle}{\emph{Advances in Neural Information Processing
  Systems}}, Vol.~\bibinfo{volume}{30}. \bibinfo{publisher}{Curran Associates,
  Inc.}
\newblock
\urldef\tempurl%
\url{https://proceedings.neurips.cc/paper/2017/file/3f5ee243547dee91fbd053c1c4a845aa-Paper.pdf}
\showURL{%
\tempurl}


\bibitem[Yu et~al\mbox{.}(2022)]%
        {yu2022commercemm}
\bibfield{author}{\bibinfo{person}{Licheng Yu}, \bibinfo{person}{Jun Chen},
  \bibinfo{person}{Animesh Sinha}, \bibinfo{person}{Mengjiao~MJ Wang},
  \bibinfo{person}{Hugo Chen}, \bibinfo{person}{Tamara~L Berg}, {and}
  \bibinfo{person}{Ning Zhang}.} \bibinfo{year}{2022}\natexlab{}.
\newblock \showarticletitle{CommerceMM: Large-Scale Commerce MultiModal
  Representation Learning with Omni Retrieval}.
\newblock \bibinfo{journal}{\emph{arXiv preprint arXiv:2202.07247}}
  (\bibinfo{year}{2022}).
\newblock


\bibitem[Yu et~al\mbox{.}(2021)]%
        {yu2021mixbert}
\bibfield{author}{\bibinfo{person}{Tan Yu}, \bibinfo{person}{Xiaokang Li},
  \bibinfo{person}{Jianwen Xie}, \bibinfo{person}{Ruiyang Yin},
  \bibinfo{person}{Qing Xu}, {and} \bibinfo{person}{Ping Li}.}
  \bibinfo{year}{2021}\natexlab{}.
\newblock \showarticletitle{MixBERT for image-ad relevance scoring in
  advertising}. In \bibinfo{booktitle}{\emph{Proceedings of the 30th ACM
  International Conference on Information \& Knowledge Management}}.
  \bibinfo{pages}{3597--3602}.
\newblock


\end{thebibliography}

\end{document}